\documentclass[12pt]{article}
\usepackage{amssymb,slashed,latexsym,amsmath,graphicx}
\newsavebox{\uuunit}
\sbox{\uuunit}
    {\setlength{\unitlength}{0.825em}
     \begin{picture}(0.6,0.7)
        \thinlines
        \put(0,0){\line(1,0){0.5}}
        \put(0.15,0){\line(0,1){0.7}}
        \put(0.35,0){\line(0,1){0.8}}
       \multiput(0.3,0.8)(-0.04,-0.02){12}{\rule{0.5pt}{0.5pt}}
     \end {picture}}

\renewcommand{\theequation}{\thesection.\arabic{equation}}
\csname @addtoreset\endcsname{equation}{section}

\long\def\symbolfootnote[#1]#2{\begingroup%
\def\thefootnote{\fnsymbol{footnote}}\footnote[#1]{#2}\endgroup}

\begin{document}

\begin{titlepage}
\begin{flushright}
CERN-PH-TH/2011-200\\
August 15, 2011\\
\end{flushright}
\vspace{.5cm}
\begin{center}
\baselineskip=16pt {\LARGE Higgs Decay $H\rightarrow\gamma\gamma$ through a $W$ Loop:\\
\vspace{12pt}
Difficulty with Dimensional Regularization}\\
\vfill
{\large R.~Gastmans$^{1,}$\symbolfootnote[1]{Work supported in part
by the FWO-Vlaanderen, project G.0651.11, and in part by the Federal
Office for Scientific, Technical and Cultural Affairs through the
`Interuniversity Attraction Poles Programme -- Belgian Science
Policy' P6/11-P.}}, {\large Sau Lan Wu$^{2,}$\symbolfootnote[2]{Work
supported in part by the United States Department of Energy Grant
No.~DE-FG02-95ER40896.}},
and {\large Tai Tsun Wu$^3$, 
  } \\
\vfill
{\small $^1$ Instituut voor Theoretische Fysica, Katholieke Universiteit Leuven,\\
       Celestijnenlaan 200D, B-3001 Leuven, Belgium. \\  \vspace{6pt}
$^2$ Department of Physics, University of Wisconsin, Madison WI 53706, USA.\\
\vspace{6pt} $^3$ Gordon McKay Laboratory, Harvard University,
Cambridge MA 02138, USA,\\ \vspace{6pt}
       and\\ \vspace{6pt}
Theory Division, CERN, CH-1211 Geneva 23, Switzerland.   \\[2mm] }
\end{center}
\vfill
\begin{center}
{\bf Abstract}
\end{center}
{\small Since the photon has no mass, it does not couple directly to
the Higgs particle. This implies that the one-loop correction to the
decay~$H\rightarrow\gamma\gamma$ is necessarily finite. Therefore,
this correction should be calculable without introducing either
regularization or ghosts. Such a calculation is carried out in this
paper for the case of one~$W$ loop. The result obtained this way
turns out {\it not} to agree with the previous, well-known one, and
it is argued that the present result is to be preferred because it
satisfies the decoupling theorem. The discrepancy can be traced to
the use of dimensional regularization in the previous result.}

\end{titlepage}
\addtocounter{page}{1}
\newpage
\indent 1. From the experimental data of the Large Electron-Positron
colliding accelerator (LEP) at CERN~\cite{R1,R2}, there was first
possible evidence for the Higgs particle~\cite{R3} at a mass of
about~115~GeV$/c^2$. It will be important for the Large Hadron
Collider (LHC) to either confirm or contradict this first possible
evidence.

If the Higgs particle has indeed a mass around this value, then, at
the LHC, one of the good ways to detect this particle is through the
decay
\renewcommand{\theequation}{\arabic{equation}}
\begin{equation}
H\rightarrow\gamma\gamma\,\label{eq1}
\end{equation}
because experimentally this decay mode can be seen cleanly.

Within the standard model of Glashow, Weinberg, and Salam~\cite{R4},
since the Higgs particle couples most strongly to heavy particles,
this decay~\eqref{eq1} proceeds predominantly through a top loop and
a $W$ loop. The contribution from one top loop was first obtained by
Rizzo~\cite{R5}; it is the purpose of the present paper to study the
contribution from one $W$ loop.

This decay~\eqref{eq1} through one $W$ loop was already studied many
years ago~\cite{R6,R7,R8}. A different point of view is taken here,
and our result does not agree with the earlier one. A
comparison of the results and the reason for discrepancy
is to be given in Sec.~9\\

2. In the Lagrangian of the standard model~\cite{R4}, there is no
coupling of the Higgs particle to the photons. It therefore follows
that the one-loop contribution to the decay~\eqref{eq1} {\it must}
be finite. This fact then implies that no regularization of any sort
is necessary for this calculation; in particular, dimensional
regularization is not needed.

In the present paper, we take the attitude that concepts that are
not necessary for the present calculation are to be avoided. As a
first application of this point of view, throughout the present
calculation,
\begin{equation}
\mbox{the space-time dimension} = 4\,.\label{eq2}
\end{equation}
In particular, there will be no non-integer dimensions.

Let this point of view be take one step further. The perturbative
calculation for the decay~\eqref{eq1} through one $W$ loop is to be
carried out in the most straightforward way. What this means is that
the present work is to be carried out in the {\it unitary gauge}.

This point requires some discussion. It is known that the unitary
gauge has many desirable properties, but it has the major handicap
of very serious difficulties with renormalization. Why? On the one
hand, because of gauge invariance, the sum of the contributions from
the various diagrams, properly interpreted, does not depend on the
gauge, and is, in particular, the same for the unitary gauge and the
renormalizable gauges, provided all the external lines are on mass
shell. On the other hand, for the purpose of renormalization, the
various diagrams cannot be treated together. It is the separate
treatment of the various diagrams that makes the unitary gauge
unsuitable for renormalization.

For the present problem of the $W$ loop contribution to the Higgs
decay~\eqref{eq1}, the amplitude, as discussed above, is convergent,
and hence there is no need to be concerned with renormalization.
Accordingly, there is no fundamental difficulty to carry out the
theoretical considerations in the unitary gauge.\\

3. We proceed to calculate, in the unitary gauge, the
decay~\eqref{eq1} through one~$W$ loop. Since calculations in the
unitary gauge are not entirely familiar to everybody, the two
salient features are to be reviewed in the present section.

The first salient feature is that, in the unitary gauge, there is no
ghost of any sort; there are only the physical particles. For the
present problem, since the only internal particle is the~$W$, the
relevant Feynman rules consist of the $W$ propagator, the~$HWW$
vertex, the~$WW\gamma$ vertex, and the~$WW\gamma\gamma$ vertex;
the~$W$ propagator is shown in Fig.~\ref{fig1}, the vertices being
the same in all gauges. For the decay~\eqref{eq1} through
\begin{figure}[t!]
\refstepcounter{figure} \label{fig1} \addtocounter{figure}{-1}
\includegraphics[width=.9\textwidth]{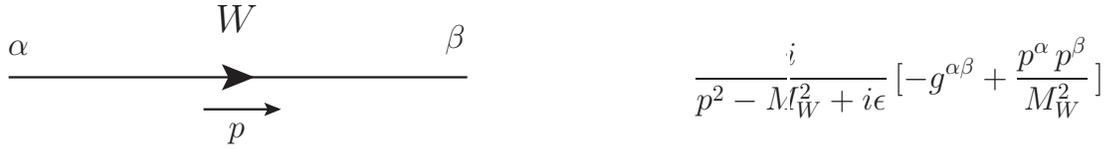}
\caption{The Feynman rule for the~$W$ propagator in the unitary
gauge with~$M_W$ the $W$ mass.}
\end{figure}
one~$W$ loop, the Feynman rules lead to only three diagrams, which
are shown in Fig.~\ref{fig2}.
\begin{figure}[t!]
\refstepcounter{figure} \label{fig2} \addtocounter{figure}{-1}
\begin{center}
\includegraphics[width=1.\textwidth]{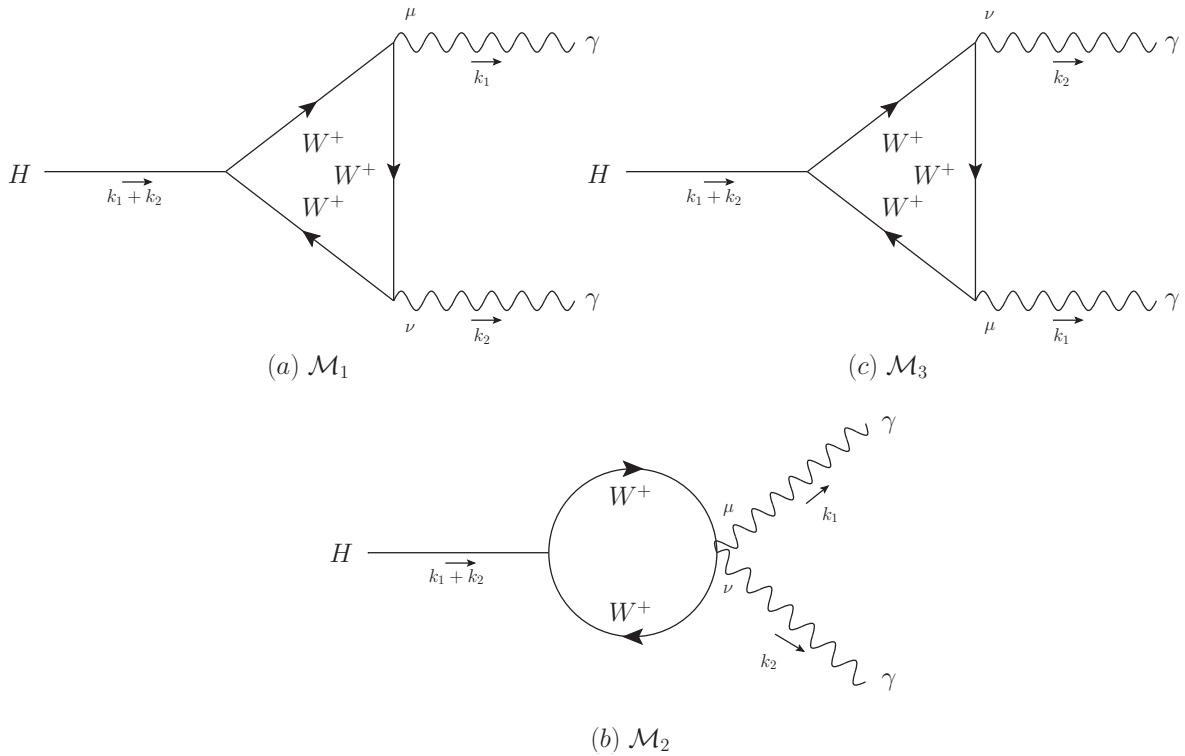}
\caption{The three diagrams in the unitary gauge for the decay
$H\rightarrow\gamma\gamma$ via a $W$ loop.}
\end{center}
\end{figure}

The second salient feature is more complicated. Because of the~$W$
propagator of Fig.~\ref{fig1}, each of the three diagrams of
Fig.~\ref{fig2} is highly divergent, in fact quartically divergent.
For any integral that is linearly divergent or worse, it is not
allowed to shift the variable of integration. Since the quantity of
interest is the sum of the contributions from the three diagrams of
Fig.~\ref{fig2}, this means that the choices of the momentum
variables for the three diagrams are inter-related. Therefore, in
order to use the unitary gauge, it is necessary to solve first the
non-trivial problem of the proper choice of the momentum variable
for the three diagrams of Fig.~\ref{fig2}.

Issues of choosing the momentum variable are well known in quantum
field theory.

(a) The best known case occurs in the context of the Ward identity
in quantum electrodynamics~\cite{R9}. This Ward identity gives a
relation between the electron self-energy and the~$ee\gamma$ vertex,
and plays an important role in disentangling overlap divergences.
The diagrammatic verification of the Ward identity requires that the
momentum to be differentiated is taken along the electron line,
\mbox{i.e.}, the external momentum of the electron self-energy is
routed through the diagram following the electron line.

(b) In order to disentangle overlap divergences in quantum
electrodynamics, it is necessary to treat the photon self-energy in
a similar way. For the photon self-energy, there is no longer a
similar obvious routing for the external photon momentum. This
problem was solved by Mills and Yang~\cite{R10}: instead of a unique
routing of the external photon momentum, the choice of the routing
for one diagram constrains the allowed routing for other diagrams.
The situation for the photon self-energy in quantum electrodynamics
is therefore similar to that of the present problem, where the
routings of the external momenta are inter-related for the three
diagrams of Fig.~\ref{fig2}.

(c) Such problems of the routings of the external momenta are in no
way limited to quantum electrodynamics. For example, they are also
present for the scalar~$\phi^4$ theory~\cite{R11}. In fact, for the
present problem, there is actually more similarity to this~$\phi^4$
theory.\\

\indent 4. Let the above general considerations be applied to the
present specific problem of calculating the matrix element for the
decay~\eqref{eq1} through a~$W$ loop. In other words, the momenta
for the internal lines of the three diagrams of Fig.~\ref{fig2} are
to be specified using the unitary gauge.

The starting point consists of the following four obvious but
important observations.

(a) The diagrams of \mbox{Fig.~\ref{fig2}(a)} and
\mbox{Fig.~\ref{fig2}(c)} can be obtained from each other by
exchanging the two external photon lines. It is therefore
sufficient, for the present purpose, to concentrate first on, say,
the diagram of \mbox{Fig.~\ref{fig2}(a)}.

(b) Under the same exchange of the external photon lines, the
diagram of \mbox{Fig.~\ref{fig2}(b)} remains unchanged.

(c) The diagram of \mbox{Fig.~\ref{fig2}(b)} can be obtained from
that of \mbox{Fig.~\ref{fig2}(a)} by ``shrinking'' the vertical~$W$
line connecting the two external photon vertices.

(d) Let $k$ be the~$W$ loop momentum to be integrated over, then the
sign of this~$k$, \mbox{i.e.}, whether~$k$ or~$-k$ is used, is
arbitrary.

Because of (b), it is convenient, for the study of the momentum
assignments in the unitary gauge, to begin with the diagram of
\mbox{Fig.~\ref{fig2}(b)}. In this case, the external momenta~$k_1$
and~$k_2$ must be routed symmetrically between the two internal~$W$
lines, thus leading to the routing of Fig.~\ref{fig3}. In this way,
\begin{figure}[t!]
\refstepcounter{figure} \label{fig3} \addtocounter{figure}{-1}
\begin{center}
\includegraphics[width=.8\textwidth]{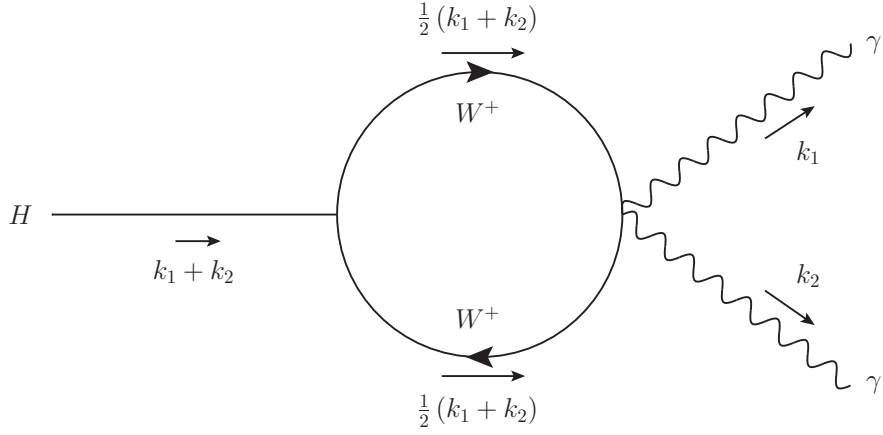}
\caption{The routing of the external momenta~$k_1$ and~$k_2$ for the
diagram of \mbox{Fig.~\ref{fig2}(b)}.}
\end{center}
\end{figure}
the momentum assignment of \mbox{Fig.~\ref{fig4}(b)} is obtained.
\begin{figure}[t!]
\refstepcounter{figure} \label{fig4} \addtocounter{figure}{-1}
\begin{center}
\includegraphics[width=1.\textwidth]{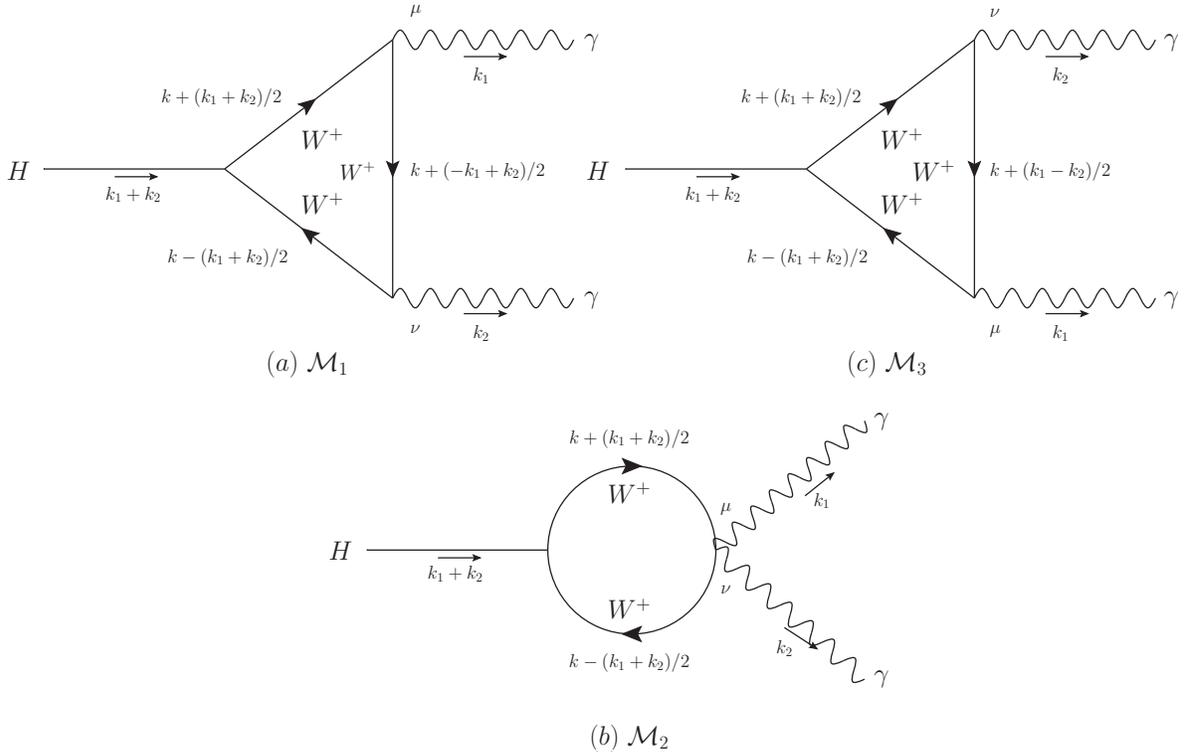}
\caption{The three diagrams in the unitary gauge for the decay
$H\rightarrow\gamma\gamma$ via a $W$ loop.}
\end{center}
\end{figure}
Note that there is nothing that specifies whether~$k$ or~$-k$ should
be used; see~(d) above.

Because of (c), it is natural to route the external photon momenta
so that the upper internal~$W$ line of \mbox{Fig.~\ref{fig2}(a)} and
\mbox{Fig.~\ref{fig2}(b)} carry the same momentum. This leads to the
momentum assignment of the upper internal~$W$ line as shown in
\mbox{Fig.~\ref{fig4}(a)}. By momentum conservation, the assignment
of \mbox{Fig.~\ref{fig4}(a)} is obtained. Finally, by (a), the
assignment of \mbox{Fig.~\ref{fig4}(c)} follows.

It is interesting to find that this momentum assignment is identical
to that used for the scalar~$\phi^4$ theory many years
ago~\cite{R11}.

Once the internal momenta are specified as shown in Fig.~\ref{fig4},
it is straightforward to write down the corresponding integrands for
the matrix elements from these three diagrams. Because of the
quartic divergences, it is not allowed to integrate them separately.
Rather, these integrands have to be added together first and then
integrated.

Let the integrands for these three diagrams be called~$I_1$, $I_2$,
and~$I_3$. They are to be evaluated with the external lines on mass
shell; in particular~$(k_1+k_2)^2=M_H^2$, where~$M_H$ is the Higgs
mass.

The matrix element for the
decay~\eqref{eq1}~$H\rightarrow\gamma\gamma$ via a~$W$ loop is given
by
\begin{equation}
{\mathcal M}=ie^2g\,M_W\int\dfrac{d^4k}{(2\pi)^4}\,I\,,\label{eq11}
\end{equation}
with
\begin{equation}
 I=I_1+I_2+I_3+I_4\,,\label{eq12}
\end{equation}
where
\begin{equation}
I_4=-(I_1+I_2+I_3)\ \hbox{\vrule height 12 pt depth 10 pt}_{\
k_1=k_2=0}\,,\label{eq10}
\end{equation}
is the Dyson subtraction~\cite{R12}. In eq.~\eqref{eq11}, $e$ is the
electric charge and $g$ the SU(2) electroweak coupling constant.\\

\indent 5. As seen from Fig.~\ref{fig1}, the~$W$ propagator consists
of two terms. For large momenta~$p$, the first term is of the
order~$p^{-2}$, while the second term is of the order of~$p^0$. This
behavior of the second term is a feature of the unitary gauge, and
is the underlying reason why the contributions of each of the three
diagrams in Fig.~\ref{fig2} are separately so highly divergent, as
already discussed in Sec.~3.

Roughly speaking, the more times the second term in the~$W$
propagator is used, the more divergent is the~$k$ integration of
this term. Such highly divergent terms are expected to be canceled
if the different~$I$'s are summed first, as indicated by
eq.~\eqref{eq12}. For the diagrams of \mbox{Fig.~\ref{fig4}(a)} and
\mbox{Fig.~\ref{fig4}(c)}, this second term may be used three times
because there are three~$W$ propagators. However, these terms are
actually zero.

Next, for each of the three diagrams of Fig.~\ref{fig4}, this second
term of the~$W$ propagator may be used twice. In the cases of
\mbox{Fig.~\ref{fig4}(a)} and \mbox{Fig.~\ref{fig4}(c)}, the two~$W$
propagators so chosen must be those connected to the external Higgs
vertex; otherwise, the contribution is again zero. These terms, with
the second term used in two~$W$ propagators, may be conveniently
referred to as the~$M_W^{-4}$ terms. These~$M_W^{-4}$ terms from the
three diagrams of Fig.~\ref{fig4} cancel each other, provided that
the various internal momenta are chosen as indicated in this
Fig.~\ref{fig4}. See especially the consequence of this point~(c) as
discussed in Sec.~4. With these~$M_W^{-6}$ (which is 0) and
the~$M_W^{-4}$ terms taken care of, the sum~$I_1+I_2+I_3$ can be
written in the following form:
\begin{equation}
I_1+I_2+I_3=I'_1+I'_2+I'_3\,,\label{eq14}
\end{equation}
where
\begin{equation}
I'_2=-3\,g_{\mu\nu}\,M_W^2\,\Big(\dfrac{1}{D_1\,D_2\,D_3}+\dfrac{1}{D_1\,D'_2\,D_3}
\Big)\label{eq15}
\end{equation}
and where the quantities~$D$ are the denominators of the~$W$
propagators.

In eq.~\eqref{eq14}, the quantities~$I'_1$ and~$I'_3$ are related by
\begin{equation}
I'_3=I'_1\ \hbox{\vrule height 12 pt depth 10 pt}_{\
k_1\leftrightarrow k_2\,,\,\mu\leftrightarrow\nu}\,;\label{eq17}
\addtocounter{equation}{1}
\end{equation}
it is convenient to separate~$I'_1$ into~$M_W^{-2}$ and $M_W^0$ terms as follows:
\begin{equation}
I'_1=I'^{(2)}+I'^{(0)}\,,\label{eq18}
\end{equation}
where
\begin{eqnarray}
\lefteqn{I'^{(2)}=\dfrac{1}{M_W^2\,D_1\,D_2\,D_3}}\nonumber\\
&&\nonumber\\
&&\times\,\Big\{\,g_{\mu\nu}\,\Big[
-\big((k+\dfrac{k_1+k_2}{2})\cdot(k-\dfrac{k_1+k_2}{2})\big)\,
(k-\dfrac{k_1-k_2}{2})^2\nonumber\\
&&\nonumber\\
&&\hspace{8cm}+(k+\dfrac{k_1+k_2}{2})^2\,(k-\dfrac{k_1+k_2}{2})^2\,\Big]\nonumber\\
&&\nonumber\\
&&+2\,\Big[-(k+\dfrac{k_1+k_2}{2})^2\,(k-\dfrac{k_2}{2})_\mu\,(k-\dfrac{k_1}{2})_\nu
-(k-\dfrac{k_1+k_2}{2})^2\,(k+\dfrac{k_2}{2})_\mu\,(k+\dfrac{k_1}{2})_\nu\nonumber\\
&&\nonumber\\
&&+4\,\big((k+\dfrac{k_1+k_2}{2})\cdot(k-\dfrac{k_1+k_2}{2})\big)\,
(k+\dfrac{k_2}{2})_\mu\,(k-\dfrac{k_1}{2})_\nu\,\Big]\,\Big\}\label{eq19}
\end{eqnarray}
and~$I'^{(0)}$ consists of terms without a factor of~$M_W^2$ in the
denominator.

Even though the right-hand side of eq.~\eqref{eq19} contains an
overall factor of~$M_W^{-2}$ while~$I'^{(0)}$ does not, the
splitting as given by eq.~\eqref{eq18} for~$I'_1$  is not quite what
is needed. The underlying reason for this complication is that the
integral
\begin{equation}
\int d^4k\,I'^{(2)}\label{eq21}
\end{equation}
is linearly, not logarithmically, divergent. This point is to be
discussed in the next section, Sec.~6.\\

\indent 6. Since linearly divergent integrals such as~\eqref{eq21}
are tricky to deal with, it is highly desirable to rewrite
the~$I'^{(2)}$ of eq.~\eqref{eq19} in such a way that the
integral~\eqref{eq21} becomes logarithmically divergent. The basic
idea of such rewriting is to average the integral under the change
of sign~$k\rightarrow -k$; see~(d) of Sec.~4.

From eq.~\eqref{eq19}, the quantity~$I'^{(2)}\,M_W^2\,D_1\,D_2\,D_3$
contains the term cubic in~$k$
\begin{equation}
\big[\,\big(k+\dfrac{k_1+k_2}{2}\big)\cdot\big(k-\dfrac{k_1+k_2}{2}\big)\,\big]\,\big
\{\,g_{\mu\nu}\,[\,k\cdot(k_1-k_2)\,]+2\,(\,k_{2\mu}\,k_\nu-k_\mu\,k_{1\nu}\,)\,\big\}\,;\label{eq22}
\end{equation}
this is the term responsible for the linear divergence mentioned
above in Sec.~5.

The first factor in the bracket can be written as the sum of three
terms: (a)~$D_2=(k-(k_1-k_2)/2)^2-M_W^2$; (b)~$M_W^2$; and
(c)~$(k\cdot(k_1-k_2))$. Of these three terms, (a) integrates to
zero because the cancelation of the factor~$D_2$ makes the integrand
odd under~$k\leftrightarrow-k$; (b) has an overall factor of~$M_W^2$
and is to be combined with the~$I'^{(0)}$ terms, this combination
being called~$I^{(0)}$, while (c) remains, leading to a
logarithmically divergent integral. The result is therefore
\begin{equation}
I'_1=I^{(2)}+I^{(0)}+\mbox{terms that are odd under
}k\rightarrow-k\,,\label{eq27}
\end{equation}
where
\begin{eqnarray}
\lefteqn{I^{(2)}=\dfrac{1}{M_W^2\,D_1\,D_2\,D_3}\,\big\{\,g_{\mu\nu}\,[\,2\,k^2\,(k_1\cdot
k_2)-4\,(k\cdot k_1)\,(k\cdot k_2)\,]}\nonumber\\
&&\nonumber\\
&&+2\,[-2\,(k_1\cdot k_2)\,k_\mu\,k_\nu+2\,(k\cdot
k_2)\,k_\mu\,k_{1\nu}+2\,(k\cdot
k_1)\,k_{2\mu}\,k_\nu-k^2\,k_{2\mu}\,k_{1\nu}\,]\,\big\}\label{eq28}
\end{eqnarray}
and
\begin{eqnarray}
I^{(0)}&=&\dfrac{1}{D_1\,D_2\,D_3}\,\big\{\,g_{\mu\nu}\,[\,6\,k_1\cdot
k_2+3\,(k-\dfrac{k_1-k_2}{2})^2\,]\nonumber\\
&&\nonumber\\
&&+\,[\,-12\,(k+\dfrac{k_2}{2})_\mu\,(k-\dfrac{k_1}{2})_\nu-6\,k_{2\mu}\,k_{1\nu}\,]\,\big\}\,.\label{eq29}
\end{eqnarray}\\

\indent 7. Consider first the quantity~$I^{(2)}$ as given by
eq.~\eqref{eq28}. Since this~$I^{(2)}$ is to be integrated
over~$d^4k$, it is convenient to use the Feynman parameters,
called~$\alpha_1$, $\alpha_2$, and~$\alpha_3$ for the three
denominators~$D_1$, $D_2$, and~$D_3$ respectively. Thus, the
appropriate change of variables is
\begin{equation}
\ell=k+\dfrac{1}{2}\,[\,(\alpha_1-\alpha_2-\alpha_3)\,k_1+(\alpha_1+\alpha_2-\alpha_3)\,k_2\,]\,.\label{eq30}
\end{equation}

From eq.~\eqref{eq28}, the numerator for~$I^{(2)}$ is
\[I^{(2)}\,M_W^2\,D_1\,D_2\,D_3\,,\]
which should first be expressed
as a polynomial in~$\ell$, and then only the part even in~$\ell$ is
kept. This even part is given as
\begin{eqnarray}
\lefteqn{g_{\mu\nu}\,[\,2\,\ell^2\,(k_1\cdot k_2)-4\,(\ell\cdot
k_1)\,(\ell\cdot k_2)\,]}\nonumber\\
&&\nonumber\\
&&+2\,[\,-2\,(k_1\cdot k_2)\,\ell_\mu\,\ell_\nu+2\,(\ell\cdot
k_2)\,\ell_\mu\,k_{1\nu}+2\,(\ell\cdot
k_1)\,k_{2\mu}\,\ell_\nu-\ell^2\,k_{2\mu}\,k_{1\nu}\,]\,.\label{eq31}
\end{eqnarray}
It is interesting to note that this expression is independent of the
Feynman parameters~$\alpha_1$, $\alpha_2$, and~$\alpha_3$.

At this point, symmetric integration may be applied so that
\begin{equation}
\ell_\alpha\,\ell_\beta\rightarrow\dfrac{1}{4}\,g_{\alpha\beta}\,\ell^2\,.\label{eq32}
\end{equation}
It should be emphasized that the factor on he right-hand side is
${1\over4}$, because the entire calculation is carried out in four
dimensions, as expressed explicitly by eq.~\eqref{eq2}.

Application of this~\eqref{eq32} for symmetric integration then
leads to the nice result that
\begin{equation}
\mbox{the numerator }\eqref{eq31}\rightarrow0\,.\label{eq33}
\end{equation}
In other words, the~$I^{(2)}$ as given by eq.~\eqref{eq28} gives
zero when integrated over~$d^4k$.

The consequence of this simplification is that, of the various terms
on the right-hand side of eq.~\eqref{eq27}, only one --- $I^{(0)}$
--- gives a non-zero contribution when integrated over~$d^4k$.\\

\indent 8. The rest of the calculation is completely straightforward
although still somewhat tedious. The result of the present
considerations using the unitary gauge is
\begin{equation}
{\cal M}=-\dfrac{e^2g}{8\pi^2M_W}\,[\,k_{2\mu}\,k_{1\nu} -
g_{\mu\nu}\,(k_1\cdot
k_2)\,]\,[\,3\,\tau^{-1}+3\,(2\,\tau^{-1}-\tau^{-2})\,f(\tau)\,]\,,\label{eq34}
\end{equation}
where
\begin{equation}
\tau=\dfrac{M_H^2}{4\,M_W^2}\label{eq35}
\end{equation}
and
\begin{equation}
f(\tau)=\left\{\begin{array}{lcc}
[\,\sin^{-1}\sqrt{\tau}\,]^2&\mbox{for}&\tau\leq 1\,,\\[5mm]
-{\displaystyle\frac{1}{4}\,\left[\ln\frac{1+\sqrt{1-\tau^{-1}}}{1-\sqrt{1-\tau^{-1}}}-i\pi\right]^2}
&\mbox{for}&\tau>1\,.\end{array}\right.\label{eq36}
\end{equation}\\

\indent 9. The most interesting and unexpected aspect of the present
result as given by eq.~\eqref{eq34} for the matrix element of the
decay~$H\rightarrow\gamma\gamma$ through a~$W$ loop is that it
disagrees with the previous one~\cite{R6,R7,R8}. The previous result
is
\begin{equation}
{\cal M}=-\dfrac{e^2g}{8\pi^2M_W}\,[\,k_{2\mu}\,k_{1\nu} -
g_{\mu\nu}\,(k_1\cdot k_2)\,]\,[\,2
+3\,\tau^{-1}+3\,(2\,\tau^{-1}-\tau^{-2})\,f(\tau)\,]\,,\label{eq37}
\end{equation}
The present answer~\eqref{eq34} differs from this~\eqref{eq37} by
the term~2 in the second bracket. This implies that, under no
circumstance, the two answers can agree exactly.

What is the basic difference between the present derivation of our
result~\eqref{eq34} and the previous one of~\eqref{eq36}? As already
mentioned in Sec.~2, the philosophy or the point of view for the
present derivation is based on the following two related points:
\renewcommand{\labelenumi}{(\alph{enumi})}
\begin{enumerate}
\item Since the Higgs particle does not couple directly to the
massless photon, the decay~$H\rightarrow\gamma\gamma$ to one-loop
order must be finite. Under this circumstance, there should be
straightforward calculation for the present process of this Higgs
decay through a~$W$ loop.
\item For such a finite calculation, there
is no reason to introduce complications such as regularization of
any sort and/or the use of various ghost particles.
\end{enumerate}
What has been carried out in this paper is just such a
straightforward calculation. This is to be contrasted with the
derivation of the previous result~\eqref{eq37}, where both
dimensional regularization and ghosts are used~\cite{R6,R7,R8}.

When two results are derived on the basis of very different points
of view, it is often difficult to pinpoint the reason for the
different results. However, in the present case, there is one
eminently reasonable explanation for the difference, and this
difference is likely to have far-reaching consequences.

In Sec.~7, the conclusion is reached that~$\int d^4k\,I^{(2)}$ is
zero. If dimensional regularization is used, then the integral
should instead be~$\int d^n k\,I^{(2)}$. With the dimension
being~$n$ instead of~4, eq.~\eqref{eq32} must be replaced by
\begin{equation}
\ell_\alpha\,\ell_\beta\rightarrow\dfrac{1}{n}\,g_{\alpha\beta}\,\ell^2\,,\label{eq38}
\end{equation}
and, consequently,~\eqref{eq33} is instead
\begin{equation}
\mbox{the numerator of }\eqref{eq31}\rightarrow
2\,(1-\dfrac{4}{n})\,\ell^2\,[\,g_{\mu\nu}\,(k_1\cdot
k_2)-k_{2\mu}k_{1\nu}\,]\,,\label{eq39}
\end{equation}
which is not zero when~$n\neq4$.

This factor of~$(4-n)$ on the right-hand side of~\eqref{eq39} is
canceled by another factor \mbox{of~$1/(4-n)$} coming from the
$n$-dimensional integration over~$d^n k$, leaving a finite answer
instead of zero. Therefore, when dimensional regularization is used,
this~$I^{(2)}$ contributes a finite term instead of the zero of
Sec.~7. Indeed, this contribution is precisely the term~2, which is
the difference between the previous result for the
decay~$H\rightarrow\gamma\gamma$ via a~$W$ loop and the present one.

As emphasized in~(b) above, there is no justification to use
dimensional regularization for the present problem. This is the
first theoretical reason why the present result is to be preferred.
This theoretical reason is supported by the second independent one
as follows.

This is connected to the decoupling theorem~\cite{R13}. In the
present context of the Higgs decay, decoupling refers to the
phenomenon that the decay~$H\rightarrow\gamma\gamma$ becomes weaker
and weaker when the Higgs mass increases without bound. While this
``decoupling theorem'' is not really a theorem in the sense that its
validity has been established beyond doubt, it does make good
physical sense. Within the present context, it is known that the
decoupling theorem does hold for the
decay~$H\rightarrow\gamma\gamma$ through a top loop~\cite{R5}.

A major qualitative difference between~\eqref{eq34} and~\eqref{eq37}
is that the present result~\eqref{eq34} does satisfy the decoupling
theorem while the previous result~\eqref{eq37} does not. Thus there
are two independent theoretical arguments that the present
formula~\eqref{eq34} for the decay matrix element
of~~$H\rightarrow\gamma\gamma$ through
a~$W$ loop is to be preferred.\\

\indent 10. It only remains to add a few important comments.
\begin{enumerate}
\item Even though the present study concerns mostly with triangle
diagrams, there are significant differences from the anomalies, for
example the Adler-Bell-Jackiw anomaly~\cite{R14}. While the effect
of this ABJ anomaly is limited entirely to the lowest-order
diagrams, the present disagreement with the earlier result
propagates to higher orders, including diagrams with
a~$H\gamma\gamma$ vertex insertion. Thus, the present result should
not be called an anomaly.
\item In the pioneering paper of Ellis, Gaillard, and Nanopoulos~\cite{R6},
the Higgs mass was considered to be small, as generally believed at
that time. Compared with their result, in this limit, the present
result is smaller by a factor of~5/7. For larger Higgs
\mbox{mass~\cite{R6,R7,R8}}, this ratio first increases but
eventually decreases in absolute value to approach zero, consistent
with the present result satisfying the decoupling
theorem~\cite{R13}.

\item It was pointed out in Sec.~4 that, for this lowest-order
one~$W$ loop diagram, the routing of the external momenta is
identical to that used for the scalar~$\phi^4$ theory~\cite{R11}. It
is likely that this relation also holds for higher-order diagrams.
Such a relation would be very interesting because there are a number
of unanswered problems in this Ref.~\cite{R11}, and the question may
be raised whether the relation, if it indeed holds to higher orders,
is of help in studying these open problems.
\item The purpose of this paper is to give a correct and
straightforward calculation of the matrix element for the
decay~$H\rightarrow\gamma\gamma$ through one~$W$ loop. In a more
general context, the significance of the present paper is probably
in pointing out that dimensional regularization must be used with
care: continuity in the space-time dimensionality is not always
true.
\end{enumerate}

\medskip

\section*{Acknowledgments}

\noindent One of us (T.T.W.) is greatly indebted to the CERN Theory
Group for their hospitality.
\newpage


\begin{thebibliography}{99}
\bibitem{R1}
ALEPH Collaboration, R.~Barate {\it et al.}, Phys.\ Lett.\ B 495
(2000) 1; DELPHI Collaboration, P.~Abreu {\it et al.}, Phys.\ Lett.\
B 499 (2001) 23; OPAL Collaboration, G.~Abbiendi {\it et al.},
Phys.\ Lett.\ B 499 (2001) 38; L3 Collaboration, P.~Achard {\it et
al.}, Phys.\ Lett.\ B 517 (2001) 319.
\bibitem{R2}
P.A.~McNamara III and S.L.~Wu, Rep.\ Prog.\ Phys.\ 65 (2002) 465.
\bibitem{R3}
F.~Englert and R.~Brout, Phys.\ Rev.\ Lett.\ 13 (1964) 321;
P.W.~Higgs, Phys.\ Lett.\ 12 (1964) 132; G.S.~Guralnik, C.R.~Hagen,
and T.W.~Kibble, Phys.\ Rev.\ Lett.\ 13 (1964) 585.
\bibitem{R4}
S.L.~Glashow, Nucl.\ Phys.\ 22 (1961) 579; S.~Weinberg, Phys.\ Rev.\
Lett.\ 19 (1967) 1264; A.~Salam in {\it Elementary Particle Physics:
Relativistic Groups and Analyticity. Eighth Nobel Symposium},
ed.~N.~Svartholm (Stockholm, Almqvist and Wiksell, 1968), p.~367.
\bibitem{R5}
T.G.~Rizzo, Phys.\ Rev.\ D 22 (1980) 178.
\bibitem{R6}
J.R.~Ellis, M.K.~Gaillard, and D.V.~Nanopoulos, Nucl.\ Phys.\
B 106 (1976) 292.
\bibitem{R7} B.L.~Ioffe and V.A.~Khoze, Sov.\ J.\ Part.\ Nucl.\ 9 (1978) 50
[Fiz.\ Elem.\ Chast.\ Atom.\ Yadra 9 (1978) 118].
\bibitem{R8}
M.A.Shifman, A.I.Vainshtein, M.B.~Voloshin, and V.I.~Zakharov, Sov.\
J.\ Nucl.\ Phys.\ 30 (1979) 711 [Yad.\ Fiz.\ 30 (1979) 1368].
\bibitem{R9}
J.C.~Ward, Phys.\ Rev.\ 78 (1950) 182.
\bibitem{R10}
R.L.~Mills and C.N.~Yang, Prog.\ Theor.\ Phys.\ Suppl.\ 37 (1966) 507.
\bibitem{R11}
T.T.~Wu, Phys.\ Rev.\ 125 (1962) 1436.
\bibitem{R12}
F.J.~Dyson, Phys.\ Rev.\ 75 (1949) 486, 1736.
\bibitem{R13}
T.~Applequist and J.~Carazzone, Phys.\ Rev.\ D 11 (1975) 2856.
\bibitem{R14}
S.~Adler, Phys.\ Rev.\ 177 (1969) 2426; J.S.~Bell and R.~Jackiw,
Nuovo Cimento A 60 (1969) 47.






\end{thebibliography}
\end{document}